\documentclass[fleqn,usenatbib]{mnras}

\usepackage{newtxtext,newtxmath}

\usepackage[T1]{fontenc}

\DeclareRobustCommand{\VAN}[3]{#2}
\let\VANthebibliography\thebibliography
\def\thebibliography{\DeclareRobustCommand{\VAN}[3]{##3}\VANthebibliography}

\usepackage{graphicx}	
\usepackage{amsmath}	
\usepackage[style=base,labelformat=simple]{subcaption} 

\newcommand{\sunxspex}{sunkit-spex}
\newcommand{\evtid}{SOL2021-11-17T21:14}
\newcommand{\orcid}[1]{\href{https://orcid.org/#1}{\textcolor[HTML]{A6CE39}{\aiOrcid}}}
\usepackage[normalem]{ulem}

\title[X-ray microflare emission and nested sampling]{Detecting non-thermal emission in a solar microflare using nested sampling}

\author[K. Cooper et al.]
{Kristopher Cooper,$^{1,2}$\thanks{E-mail: coop0502@umn.edu (KC)}
Iain G. Hannah,$^{2}$
Lindsay Glesener,$^{1}$
Brian W. Grefenstette,$^{3}$
\\
$^{1}$School of Physics \& Astronomy, University of Minnesota Twin Cities, Minneapolis, MN 55455, USA\\
$^{2}$School of Physics \& Astronomy, University of Glasgow, University Avenue, Glasgow G12 8QQ, UK\\
$^{3}$Cahill Center for Astrophysics, California Institute of Technology, 1216 East California Boulevard, Pasadena, CA 91125, USA\\
}

\date{Accepted XXX. Received YYY; in original form ZZZ}

\pubyear{2023}

\begin{document}
\label{firstpage}
\pagerange{\pageref{firstpage}--\pageref{lastpage}}
\maketitle


\begin{abstract}
Microflares are energetically smaller versions of solar flares, demonstrating the same processes of plasma heating and particle acceleration. However, it remains unclear down to what energy scales this impulsive energy release continues, which has implications for how the solar atmosphere is heated. The heating and particle acceleration in microflares can be studied through their X-ray emission, finding predominantly thermal emission at lower energies; however, at higher energies it can be difficult to distinguish whether the emission is due to hotter plasma and/or accelerated elections. We present the first application of nested sampling to solar flare X-ray spectra, an approach which provides a quantitative degree of confidence for one model over another. We analyse NuSTAR X-ray observations of a small active region microflare (A0.02 GOES/XRS class equivalent) that occurred on 2021~November~17, with a new Python package for spectral fitting, \sunxspex{}, to compute the parameter posterior distributions and the evidence of different models representing the higher energy emission as due to thermal or non-thermal sources. Calculating the Bayes factor, we show there is significantly stronger evidence for the higher energy microflare emission to be produced by non-thermal emission from flare accelerated electrons than by an additional hot thermal source. Qualitative confirmation of this non-thermal source is provided by the lack of hotter (10~MK) emission in SDO/AIA’s EUV data. The nested sampling approach used in this paper has provided clear support for non-thermal emission at the level of~3$\times$10$^{24}$~erg~s$^{-1}$ in this tiny microflare.
\end{abstract}




\section{Introduction}

Flares are explosive releases of energy that take place in active regions (ARs) thought to be produced and powered by magnetic reconnection of stressed magnetic fields in the solar atmosphere \citep{priest_magnetic_2002}. The reconnection can release energies up to $\sim$10$^{32}$~erg and readily drives mass flows, plasma heating, and particle acceleration \citep{fletcher_observational_2011, benz_flare_2017}. Flares produce emission across the electromagnetic spectrum and are commonly characterised by their GOES 1--8~\AA{} soft X-ray (SXR) emission and thermal energy release. Flares with SXR emission $<$10$^{-6}$~W~m$^{-2}$ and energies between 10$^{26}$--10$^{28}$~erg are commonly referred to as \textit{microflares} \citep{lin_solar_1984, hannah_microflares_2011}.

\begin{figure*}
    
    \begin{subfigure}{0.58\linewidth}
        \includegraphics[width=\linewidth]{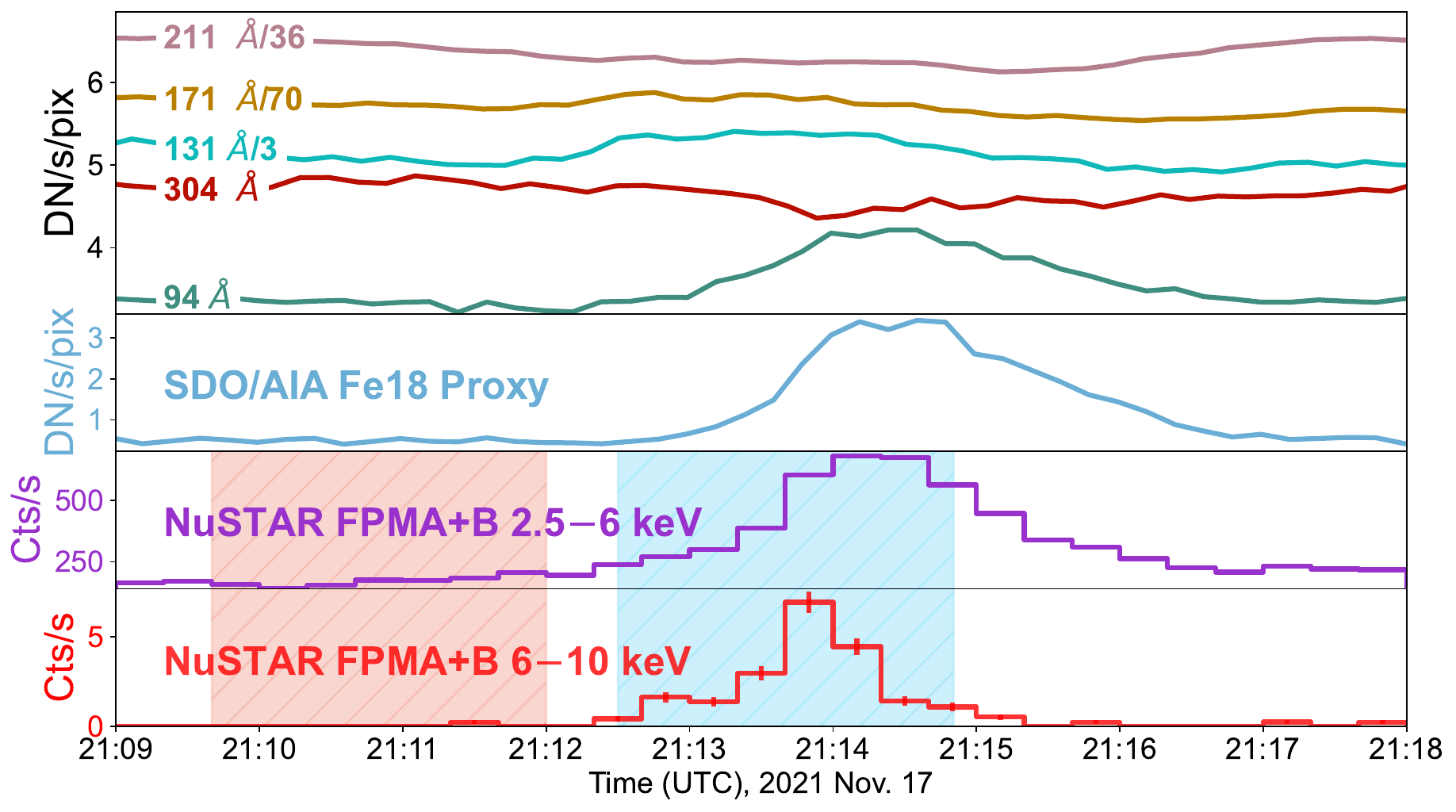}
    \end{subfigure}
    \begin{subfigure}{0.415\linewidth}
        \includegraphics[width=\columnwidth]{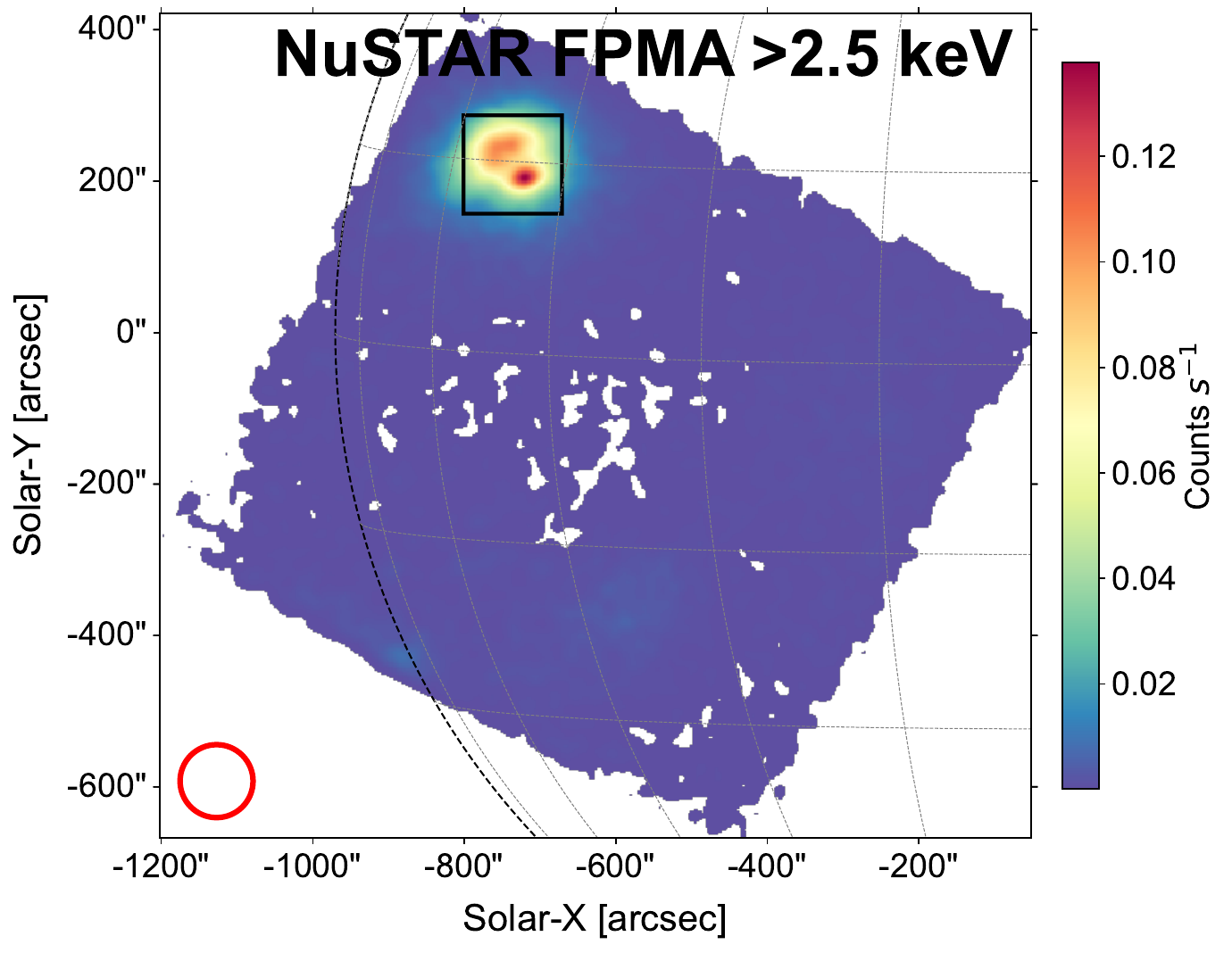}
    \end{subfigure}
    \caption{SDO/AIA, including \ion{Fe}{xviii} proxy, and NuSTAR livetime corrected light curves over the time of microflare \evtid{} (left panels). The SDO/AIA time profiles are averaged over the black box region shown in Figure~\ref{fig:overview_sdo} and a scale factor (for plotting purposes) has been applied, as shown, to several channels. The area used to create the NuSTAR light curves is shown as a black box across AR SPoCA 26190 in the NuSTAR $>$2.5~keV integrated, livetime corrected FOV (right panel). A pre-flare (21:09:40--21:12:00~UT) and microflare (21:12:30--21:14:50~UT) time is indicated by the shaded red and blue areas in the NuSTAR light curve panels, respectively. The red circle with a radius of~48\arcsec{} shown in the right panel indicates the region size used to produce spectral fit profiles for the NuSTAR spectra.}
    \label{fig:overview}
\end{figure*}

Flares with energies on the order of $\sim$10$^{24}$~erg, termed \textit{nanoflares}, are proposed to occur everywhere on the solar disk and not just localised to ARs \citep{parker_nanoflares_1988}. If microflares and nanoflares occur frequently enough then they could provide the majority fraction of the energy heating the corona from the overall flare distribution; however, this relies on the same flaring mechanisms scaling down to the weaker events, such as non-thermal particle acceleration emission \citep{hudson_solar_1991}. It is often difficult to ascertain if weak, sub-A class microflares do indeed show positive evidence for these standard flare processes often observed in their larger counterparts.

Microflares have been extensively studied in X-rays with instruments such as the Reuven Ramaty High-Energy Solar Spectroscopic Imager \citep[RHESSI;][]{lin_reuven_2002}, Chandrayaan-2's Solar X-ray Monitor \citep[XSM;][]{vadawale_solar_2014,shanmugam_solar_2020}, and the Nuclear Spectroscopic Telescope ARray \citep[NuSTAR;][]{harrison_nuclear_2013}. Statistical and individual studies have observed hard X-ray emission during microflares from non-thermal electron acceleration showing evidence that microflares down to GOES class A0.1 appear to show similar processes to their brighter counterparts \citep{christe_rhessi_2008,hannah_rhessi_2008,glesener_accelerated_2020,cooper_nustar_2021}. However, as the microflares get weaker it becomes increasingly difficult to determine if this flare behaviour continues to scale or has a lower limit. Sensitivity to these weak microflares is crucial in understanding if the flare process scales across these different classifications.

NuSTAR is an astrophysical telescope capable of observing the Sun $>$2.5~keV \citep{grefenstette_first_2016,hannah_first_2016}. NuSTAR consists of two telescopes utilising \ion{Wolter-}{i} type optics to focus X-rays onto two focal plane modules (FPMA and B), each with a field of view (FOV) of 12\arcmin$\times$12\arcmin. Each FPM takes~2.5~ms to process a detected photon during which no other trigger can be recorded; the fraction of time NuSTAR spends open to detection during an observation is quantified by the livetime. Under quiescent conditions NuSTAR's livetime can reach up to~92\% \citep{paterson_first_2023}; however, the livetime drops to $<$16\% even with small microflares of A-class or smaller \citep{cooper_nustar_2021}. This limited throughput, combined with flare X-ray spectra steeply falling with increasing energy, results in a low number of NuSTAR counts at higher energies. 

NuSTAR has observed several microflares with energies from 10$^{28}$~erg down to 10$^{26}$~erg with GOES classes of B, A, and sub-A \citep{glesener_nustar_2017,wright_microflare_2017,hannah_joint_2019, cooper_nustar_2020,cooper_nustar_2021,duncan_nustar_2021} and also quiescent Sun features outside ARs of energies down to 10$^{26}$~erg \citep{kuhar_nustar_2018,paterson_first_2023}. Several of these studies have found consistency with the presence of non-thermal emission throughout the flare evolution while \citet{glesener_accelerated_2020} and \citet{cooper_nustar_2021} show convincing evidence of non-thermal emission in an A5.7 and an estimated A0.1 microflare, respectively.\footnote{A NuSTAR solar observations overview is available at \url{https://ianan.github.io/nsigh_all/}}

It becomes more difficult with weaker flares to confidently determine the presence of any high-energy emission and even harder to determine the mechanism behind the emission. In order to determine the nature of the higher energy emission observed by an instrument from much weaker microflares, a robust method needs to be utilised which can investigate the difference between different model representations of the observed data. Nested sampling analysis \citep{skilling_nested_2004,skilling_nested_2006} can be used for this purpose. Note, this analytical technique can also be applied in general cases (e.g., in larger flares) where ambiguity exists between model fits to data. 

Nested sampling is a Bayesian technique which is used to map the posterior distribution and estimate a quantity called the \textit{evidence} which is the probability of an observation given the assumption of a model representation. The evidences of different hypotheses can be then used to compare different model fits to data, providing a level of confidence for one model over another. The nested sampling algorithm also accounts for different model parameters and different numbers of model parameters used between representations. There are many implementations of nested sampling \citep{kester_bayesicfitting_2021,ashton_nested_2022} and it is used extensively in many scientific fields \citep{knuth_bayesian_2015}; however, it is not common in the solar physics community.

In this paper, we present the first use of nested sampling in the analysis of an active region NuSTAR X-ray microflare which took place on~2021 November 17 at $\sim$21:14~UT (\evtid{}). In Section~\ref{analysis}, we investigate the microflare evolution in time, space, and energy using NuSTAR and the Solar Dynamic Observatory's Atmospheric Imaging Assembly \cite[SDO/AIA][]{lemen_atmospheric_2012} and Heliospheric and Magnetic Imager \cite[SDO/HMI][]{schou_design_2012}. We then describe the nested sampling algorithm and how this can be used to determine the most likely model representation to explain a given data-set in Section~\ref{comparing}. In Section~\ref{tornt}, we then apply the nested sampling algorithm to the NuSTAR X-ray spectral analysis for the first time where we determine the most likely explanation for the microflare emission.


\section{X-ray microflare} \label{analysis}

A NuSTAR solar campaign was performed on 2021 November 17--22 where nine hour-long observations were made of active regions present on the Sun's disk. During the third observation NuSTAR observed three microflares where the second one (microflare \evtid{}) is the subject of this paper, the other flares and observations will be investigated in future papers.

\begin{figure*}
    \includegraphics[width=\linewidth]{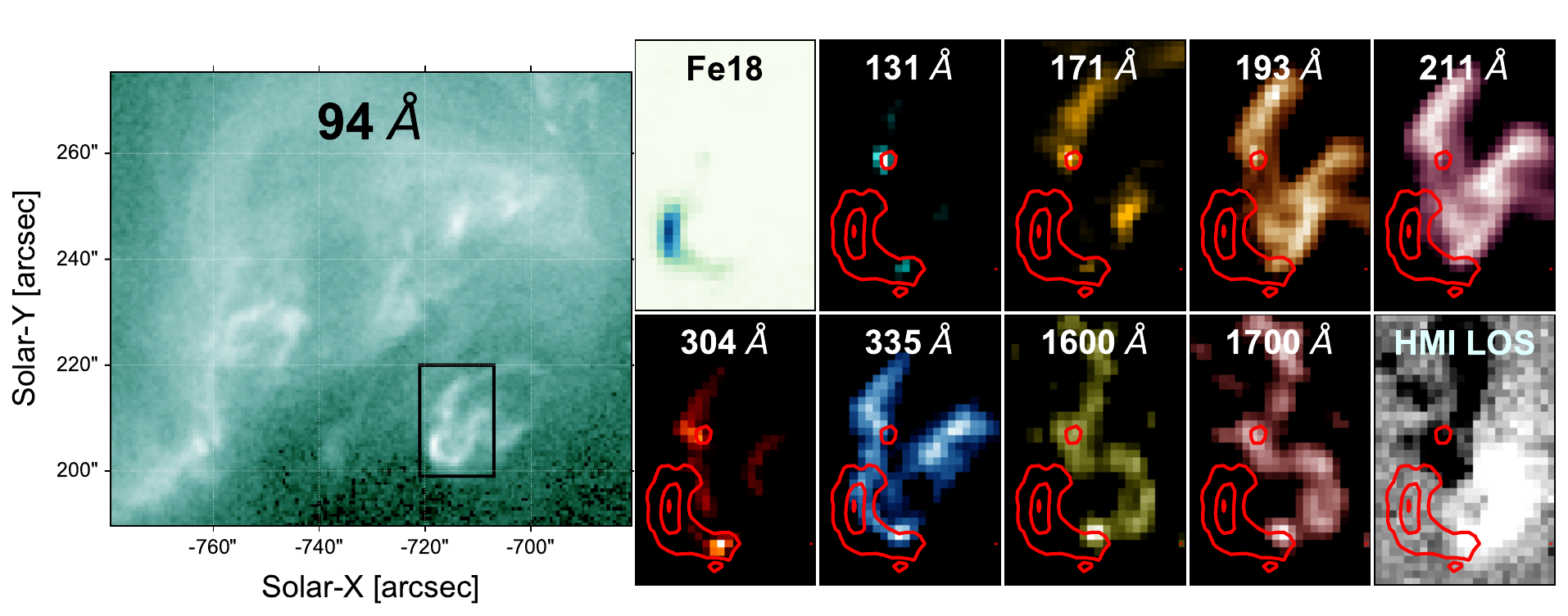}
    \caption{SDO/AIA native channels and \ion{Fe}{xviii} proxy are averaged over the microflare time (blue shaded region, Figure~\ref{fig:overview}) while the SDO/HMI line-of-sight magnetogram is taken from the start of the time range. The left panel shows the~94~\AA{} channel emission over the full active region with a FOV slightly smaller than the boxed region in the NuSTAR image from Figure~\ref{fig:overview}. A black box indicates the location of the microflare and is used to create the zoomed in images for the other emission channels (right panels). The \ion{Fe}{xviii} proxy microflaring loop emission contours at~5\%,~50\%, and~95\% of the maximum are overlaid on the zoomed panels and are shown in red.}
    \label{fig:overview_sdo}
\end{figure*}

\subsection{Time evolution} \label{analysis-te}

Microflare \evtid{} is clearly observed in X-rays with NuSTAR as shown in Figure~\ref{fig:overview}. The extreme ultraviolet (EUV) and X-ray time profiles from SDO/AIA and NuSTAR, respectively, are shown (left panels) where the SDO/AIA light curves are from a region around the microflare (see Figure~\ref{fig:overview_sdo}) and the NuSTAR time profiles are integrated over the entire AR area indicated by a black box in Figure~\ref{fig:overview} (right panel). The X-ray image (right panel) shows the microflare in the top of NuSTAR's FOV, inside the black box, as a relatively bright source compared to the more northern AR \citep[SPoCA 26190;][]{verbeeck_spoca-suite_2014} core emission.

We find that the native SDO/AIA channel light curves in Figure~\ref{fig:overview} (top of left panel) show little similarity with the X-ray time profiles (bottom of left panel) with the exception of the~94~\AA{} channel which shows a slight positive correlation and a possible negative correlation with the~304~\AA{} channel. The microflare only becomes apparent in the EUV when calculating the \ion{Fe}{xviii} proxy channel---a linear combination of the~94~\AA{},~171~\AA{}, and~211~\AA{} channels---as described in \citet{del_zanna_multi-thermal_2013}. This channel has a significant response to thermal emission from material between~4--10~MK which is a crucial range for NuSTAR microflares \citep{hannah_first_2016, hannah_joint_2019, wright_microflare_2017, glesener_accelerated_2020, cooper_nustar_2020, cooper_nustar_2021, duncan_nustar_2021}.

The \ion{Fe}{xviii} light curve shows very similar progression to the~2.5--6~keV FPMA+B NuSTAR time profile indicating this microflare heats a significant amount of material to between~4--10~MK. The higher energy X-ray range,~6--10~keV, shows a more impulsive time profile that peaks earlier, indicative of hot plasma being present in the early stages of the microflare and/or non-thermal emission from an initially accelerated electron distribution. 

A pre-flare and microflare time is defined from the NuSTAR time profiles and shown as a red and blue shaded region, respectively, in Figure~\ref{fig:overview}. The microflare time is chosen based on the more impulsive~6--10~keV NuSTAR light curve to better investigate the emission processes in the initial flaring stages.

The microflare, along with many events NuSTAR observed in the observation campaign, is obvious in the X-ray regime where it is dominant across the whole FOV; however, as indicated by the SDO/AIA light curves in Figure~\ref{fig:overview}, it is difficult to find corroborating evidence in the native EUV channels even when locating the microflare's position. Therefore, the EUV microflaring structure can only be reliably determined with information from NuSTAR and its comparison to \ion{Fe}{xviii} images.

\subsection{Spatial evolution} \label{analysis-spa}

Figure~\ref{fig:overview_sdo} shows EUV images integrated over the microflare time. Comparable in size to the black box shown in the NuSTAR FOV image in Figure~\ref{fig:overview} (right panel), the left panel of Figure~\ref{fig:overview_sdo} shows the SDO/AIA~94~\AA{} channel emission of the full AR with the microflare identified within a black box. The black box region identified to host the microflare is the area used to depicted the other panels.

The \ion{Fe}{xviii} proxy channel and, to a lesser extent, the 94~\AA{} channel show a loop structure while the other channels do not show any corresponding activity except at the apparent loop footpoints. Footpoint activity can be seen through all EUV channels; however, the footpoints do not appear to produce the same response at all levels in the solar atmosphere. The northern footpoint appears brighter in the upper transition region as shown by the relative brightness in~131~\AA{},~171~\AA{} whereas the southern footpoint shows a greater response in the chromosphere and photosphere as seen in the~304~\AA{},~1600~\AA{}, and~1700~\AA{} channels \citep{lemen_atmospheric_2012}. From the SDO/HMI panel in Figure~\ref{fig:overview_sdo} we find the loop footpoints are anchored in opposite polarity regions.

Zoomed images with NuSTAR emission only showed a single circular shape to co-align with the EUV loop structure for the~2.5--6~keV and~6--10~keV emission. The lack of distinctive shape in X-rays may be expected due to the very few high energy microflare counts and the fact the angular extent of the microflare is $\sim$10\arcsec{} which is considerably smaller than NuSTAR's angular resolution with a FWHM and HPD of~18\arcsec{} and $\sim$60\arcsec{}, respectively \citep{harrison_nuclear_2013}.

From the EUV images shown in Figure~\ref{fig:overview_sdo} we are able to obtain a microflaring loop volume by modelling it as a half-torus. We estimate an upper limit for the distance between the footpoints and the diameter of the loop to be approximately 9\arcsec{} and 2\arcsec{} (6.5$\times$10$^{8}$~cm and 1.5$\times$10$^{8}$~cm), respectively. Therefore, the volume obtained from the geometry as viewed in EUV is 1.7$\times$10$^{25}$~cm$^{3}$.

\subsection{Spectral evolution} \label{analysis-spe}

\begin{figure*}
        \includegraphics[width=0.498\linewidth]{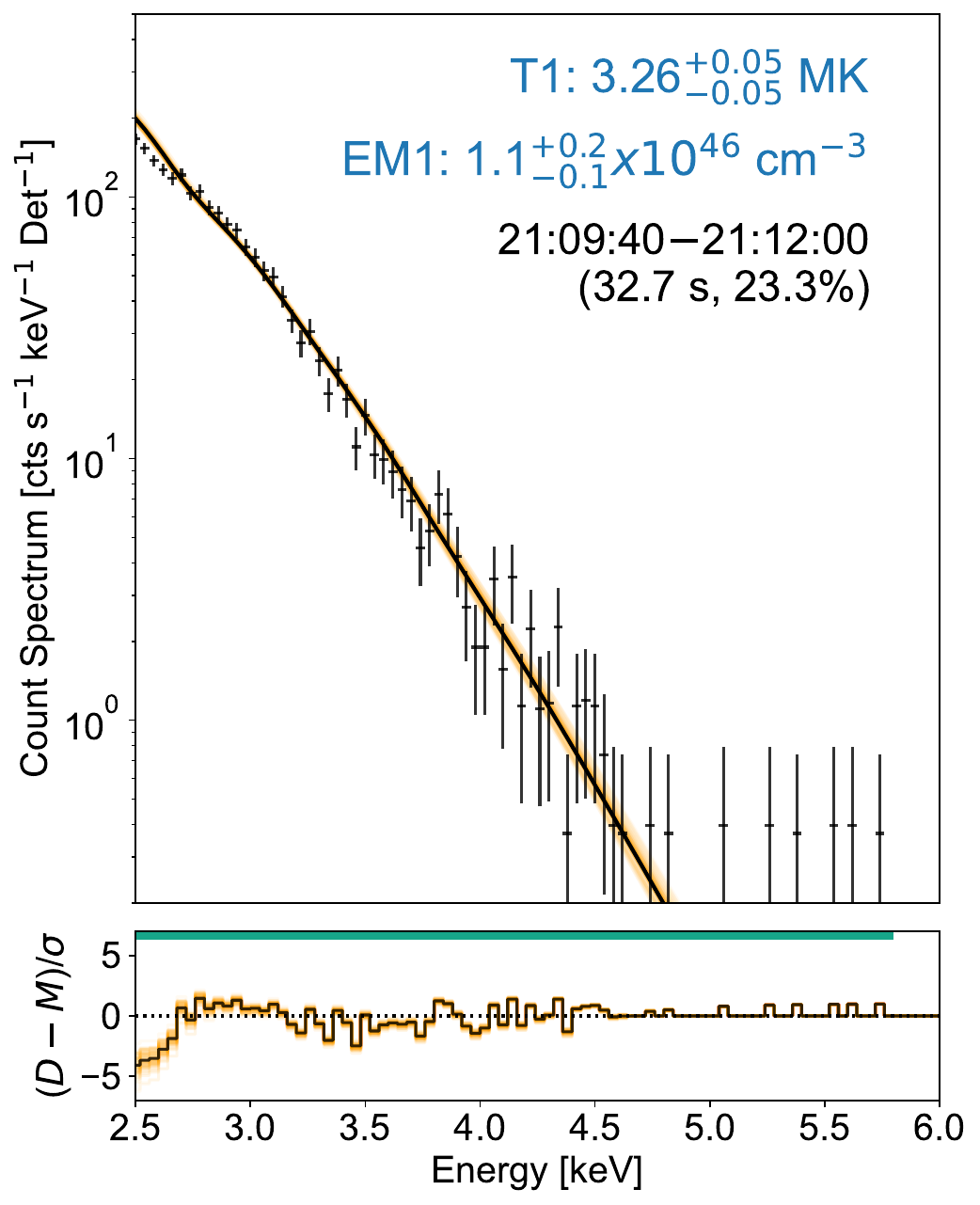}
        \includegraphics[width=0.493\linewidth]{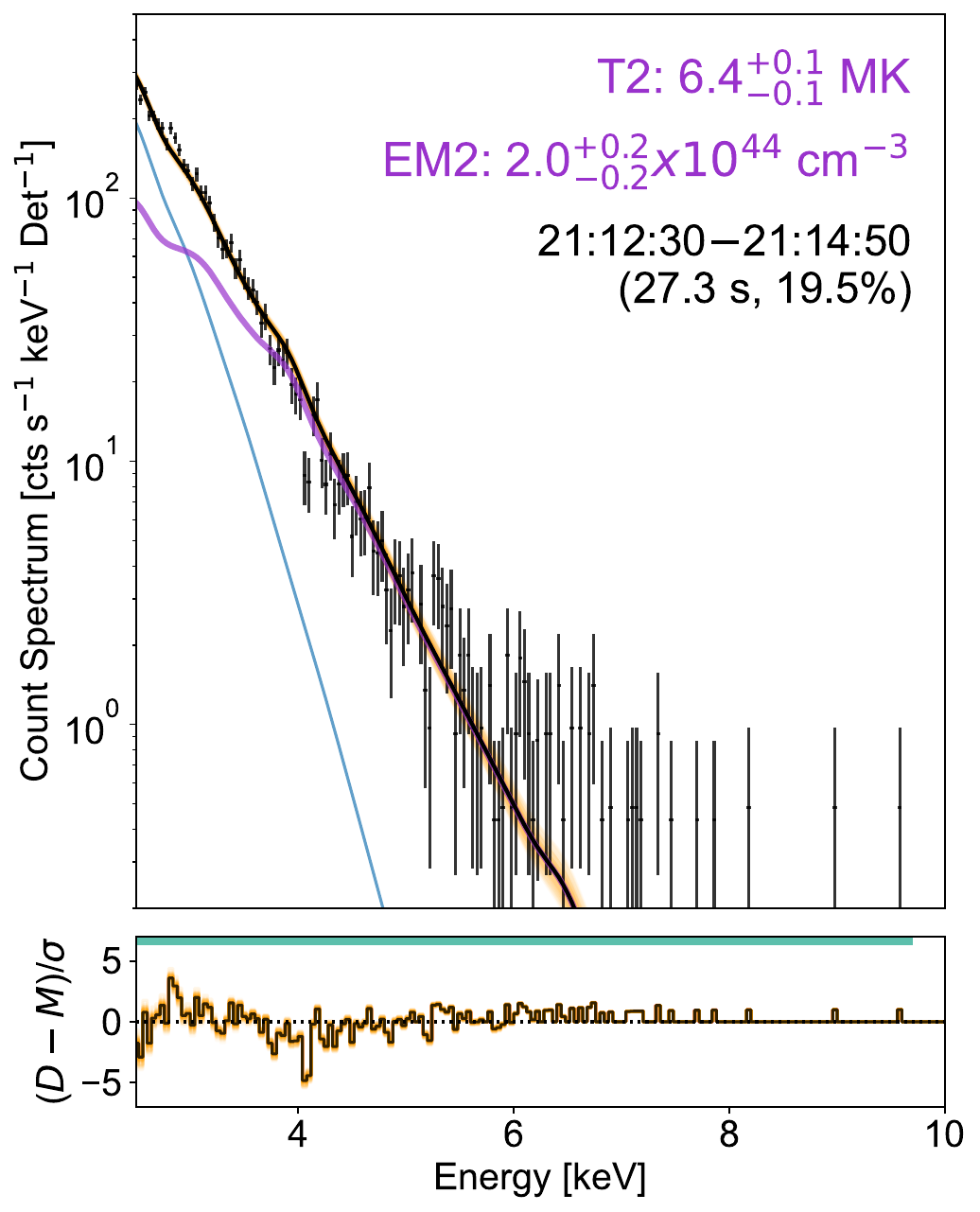}
    \caption{Simultaneous thermal model fits, using \sunxspex{}, of NuSTAR's FPMA and B emission spectra during the pre-flare (21:09:40--21:12:00~UT, left panel) and microflare (21:12:30--21:14:50~UT, right panel) times which are shown as shaded red and blue regions in Figure~\ref{fig:overview}, respectively. The pre-flare spectra are fitted with one thermal model (black) and then used as a fixed component for the microflare spectra (blue). The microflare spectra are fitted with an additional thermal model (purple) to account for the flaring emission with the overall model being shown in black. The spectra from both times are fitted over the energy ranges indicated by the horizontal green line in the residual panels. The temperatures and emission measures for each model fit are shown in their respective model colours where the error ranges are calculated from MCMC analysis and report the 1-$\sigma$ equivalent uncertainty. Additionally,~100 random model samples from the MCMC analysis are shown in orange. The relevant time ranges, effective exposures, and livetimes are also shown in each panel. The microflare fit in the right panel is also referred to as the \textit{null scenario} in relation to model comparison.}
    \label{fig:mcmc_spec}
\end{figure*}

Figure~\ref{fig:mcmc_spec} shows the spectral fitting analysis (see Section~\ref{sec:best-spec-fit}) of the NuSTAR FPMA and B grade 0 (single pixel) microflare emission on the pre-flare and microflare time using \sunxspex{}\footnote{Formerly sunxspex, a Python spectral fitting tool: \url{https://github.com/sunpy/sunkit-spex}} which utilises the Emcee package \citep{foreman-mackey_emcee_2013}. We find that the pre-flare time (Figure~\ref{fig:mcmc_spec}, left panel) is represented well with an isothermal model of temperature~3.3~MK and emission measure~1$\times$10$^{46}$~cm$^{-3}$ from the Markov chain Monte Carlo (MCMC) analysis, a common result from previous NuSTAR quiescent AR and pre-flare spectral fitting \citep{glesener_nustar_2017,wright_microflare_2017,hannah_joint_2019,cooper_nustar_2021}. The isothermal model available in \sunxspex{} is a Python version of the IDL function \verb|f_vth.pro| \citep{schwartz_rhessi_2002} and uses \verb|CHIANTI V7.1| \citep{dere_chianti_1997,landi_chiantiatomic_2013} with coronal abundances \citep{feldman_potential_1992, landi_comparison_2002}.

The pre-flare thermal model is then used as a fixed component when fitting the microflare time with an additional thermal model representing the flaring emission (Figure~\ref{fig:mcmc_spec}, right panel). This model fit provides a temperature and emission measure in a sensible range for previous NuSTAR microflares, between~4--10~MK, as discussed in Section~\ref{analysis-te}; however, it is clear that this model does not provide a satisfactory representation of the observed spectra. The residuals of the microflare fit in Figure~\ref{fig:mcmc_spec} indicate a poor fit between~2.5--6~keV and a clear count excess above the model at energies~$>$6~keV. Therefore, we find sufficient reason to include an additional model to the microflare fit in order to accurately represent the observed spectra.

Two model candidates are tested to fit the excess: a \textit{thermal scenario} which includes an additional thermal model, and a \textit{non-thermal scenario} which involves the thick-target model \citep{brown_deduction_1971,holman_implications_2011}. The thick-target model assumes a power-law distribution of electrons of spectral index $\delta$ above a low energy cut-off E$_{c}$ that lose all their energy through Coulomb collisions. These hypotheses represent either hotter plasma or non-thermal emission during the impulsive phase of the microflare. In relation to these two excess additions, the microflare fit in Figure~\ref{fig:mcmc_spec} is termed the \textit{null scenario} since the excess is not represented. Representations of the photon models utilised in each scenario are shown in Figure~\ref{fig:photon_models}. Note, unlike in the spectral analysis of brighter NuSTAR microflares, no gain correction \citep{duncan_nustar_2021} is required due to the relatively high livetime of microflare \evtid{}.

\begin{figure*}
        \includegraphics[width=0.493\linewidth]{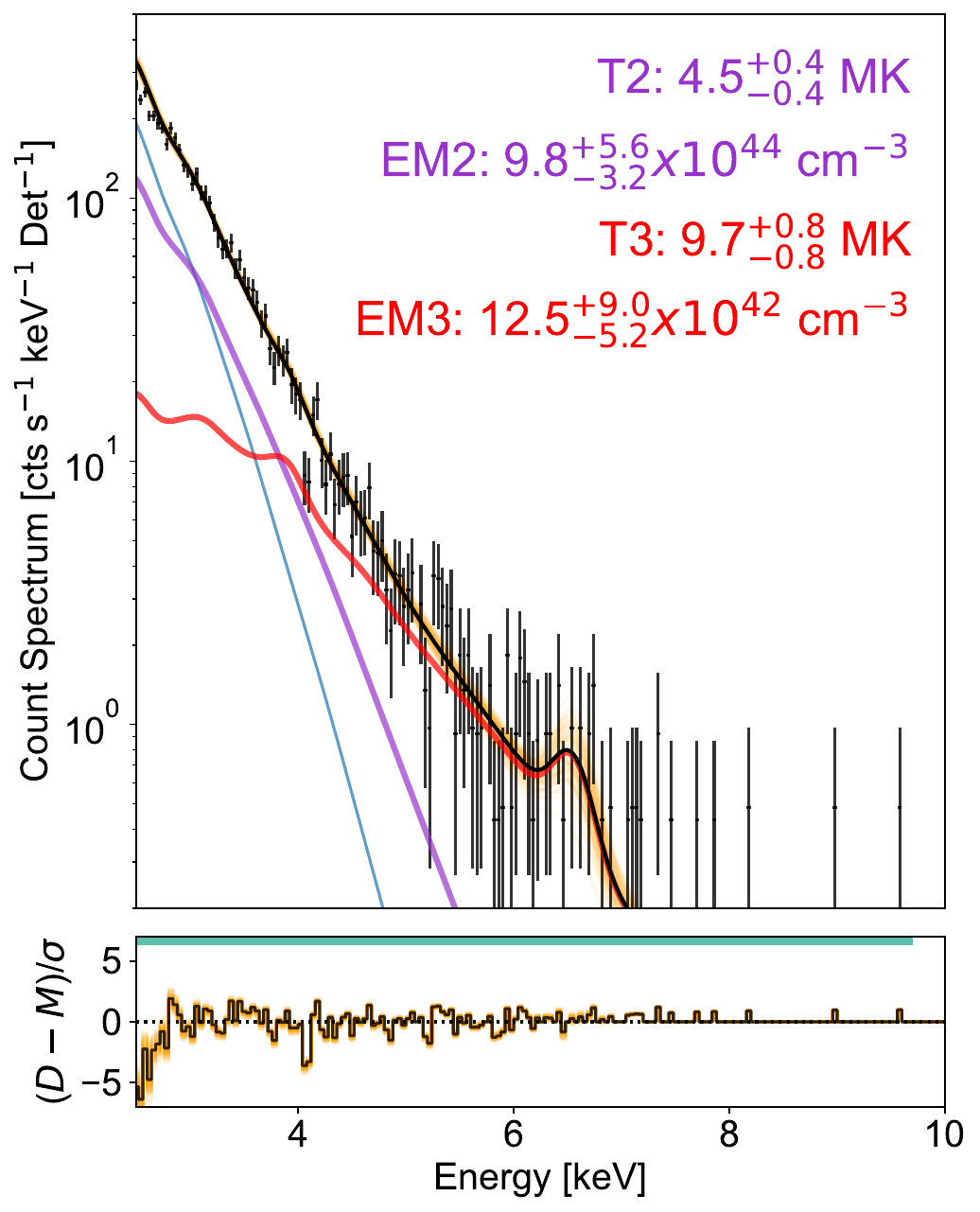}
        \includegraphics[width=0.493\linewidth]{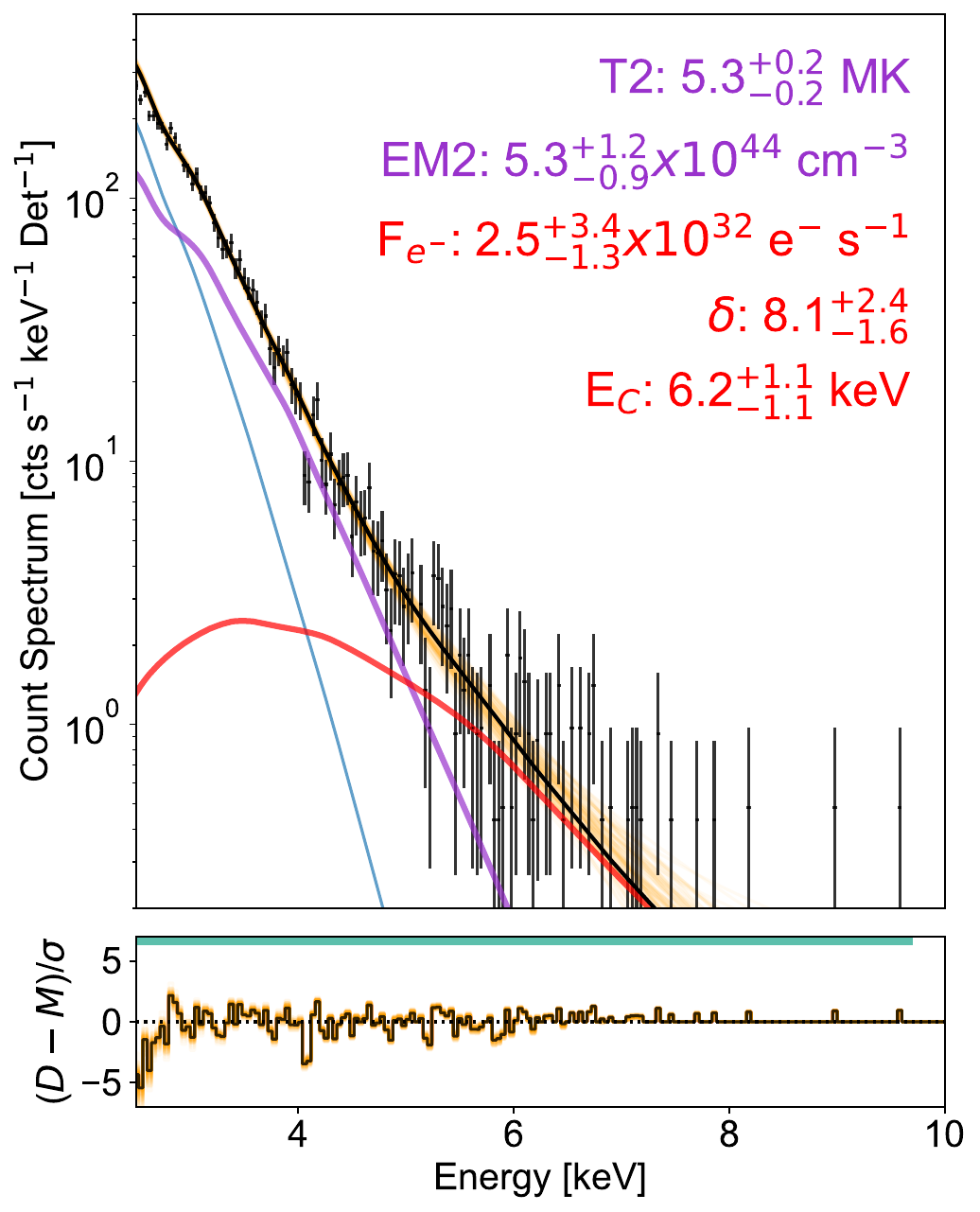}
    \caption{Two different spectral fits of the microflare spectra from the right panel of Figure~\ref{fig:mcmc_spec}, one with the count model excess fitted with a thermal model (the \textit{thermal scenario}, left panel) and the other fitted with a non-thermal model (\textit{non-thermal scenario}, right panel) shown in red. The model parameters are shown in their respective plots and colours with their error values showing 1-$\sigma$ equivalent uncertainties resulting from MCMC analysis with~100 random model samples shown in orange.}
    \label{fig:mcmc_spec_fitExcess}
\end{figure*}

Figure~\ref{fig:mcmc_spec_fitExcess} shows, in red, the thermal scenario fit (left panel) and the non-thermal scenario fit (right panel) to the microflare spectra. We find that both scenarios resolve the poor fit to the~2.5--6~keV range and represent the count excess well~$>$6~keV.

The thermally fitted excess representation proposes the presence of a relatively weak emitting source with a temperature of~$\sim$10~MK; whereas, the non-thermal case suggests thick-target emission with a $\delta$ of~$\sim$8 and an E$_{c}$ of ~$\sim$6~keV. We note the non-thermal parameters are well within those previously reported for other microflares \citep{hannah_rhessi_2008,glesener_accelerated_2020,cooper_nustar_2021}. The lower energy thermal models (purple) show a decrease in temperature and increase in emission measure compared to the values obtained in Figure~\ref{fig:mcmc_spec} (right panel).

Using the temperatures and emission measures obtained from all microflare fits, we find the microflare to be an estimated GOES class between A0.02--A0.03. The GOES class is roughly estimated using the \verb|goes_flux49.pro| IDL routine as the microflare was too weak to provide a direct GOES/XRS measurement. 

From the flare volume upper limit determined in Section~\ref{analysis-spa} and the microflare thermal parameters we find thermal energies of~1.6$^{+0.1}_{-0.1}\times$10$^{26}$~erg,~3.0$^{+0.4}_{-0.4}\times$10$^{26}$~erg, and~2.1$^{+0.1}_{-0.1}\times$10$^{26}$~erg for the null, thermal, and non-thermal scenario fits, respectively, using Equation~3 and~A4 from \citet{hannah_rhessi_2008} and \citet{aschwanden_global_2015}. In addition, the non-thermal power associated with the thick-target component in Figure~\ref{fig:mcmc_spec_fitExcess} (right panel) is calculated to be~2.7$^{+2.7}_{-1.2}\times$10$^{24}$~erg~s$^{-1}$ \citep[Equation~5 from][]{wright_microflare_2017} which results in~3.7$^{+3.8}_{-1.6}\times$10$^{26}$~erg over the course of the impulsive phase. All thermal energies are in the expected range for a weak microflare and we note in the non-thermal case both energies are consistent with each other.

Therefore, both hypotheses shown in Figure~\ref{fig:mcmc_spec_fitExcess} presents an interesting and unique conclusion, either showing a microflare of this scale producing such hot temperatures in the impulsive phase or providing positive evidence for the presence of visible non-thermal particle acceleration via a clear count excess in the spectral fits. 

\subsubsection{Best spectral fit} \label{sec:best-spec-fit}

The fits shown in Figure~\ref{fig:mcmc_spec} and~\ref{fig:mcmc_spec_fitExcess} are determined using the Poissonian likelihood, $\mathcal{L}(\theta)$, given by
\begin{equation} \label{likelihood}
    \mathcal{L}(\theta) = \prod^{N}_{i=0} \frac{M_{i}(\theta)^{D_{i}} \times e^{-M_{i}(\theta)}}{D_{i}!},
\end{equation}
where $M_{i}(\theta)$ represents the predicted model counts with parameters $\theta$ and $D_{i}$ is the observed number of counts in a given energy bin,~$i$. The number of data bins included in the optimisation is represented by $N$. 

We find the most optimum fit to the microflare spectra with the thermal, double thermal, and thermal and non-thermal models give a Poissonian log-likelihood value, $ln(\mathcal{L}_{opt})$, of~-520,~-484, and~-480, respectively. This loosely suggests the non-thermal scenario fit (Figure~\ref{fig:mcmc_spec_fitExcess}, right panel) best represents the observed spectra out of the tested models. However, it may be expected this is the case since the thick-target model has more free parameters; therefore, it is difficult to ascertain whether the non-thermal scenario should be trusted more with such a small difference in log-likelihood values given the different number of parameters available to each model.

It is possible there is a discrepancy between the thermal excess scenario (Figure~\ref{fig:mcmc_spec_fitExcess}, left panel) and the EUV time profiles displayed in Figure~\ref{fig:overview}. The double-thermal fit suggests the presence of material at $\sim$10~MK during the microflare's impulsive phase. However, the SDO/AIA~131~\AA{} channel has a significant response to material of this temperature \citep{lemen_atmospheric_2012} but the light curve does not show a noticeable increase in emission around this time. Folding the~10~MK thermal component through the EUV channel response, we find the hot component is very faint and may be too weak to produce a significant increase in~131~\AA{} channel meaning the emission mechanism of the higher X-ray energy excess remains ambiguous.

A more robust approach is needed to assess the thermal and non-thermal representations of the microflare spectra. The nested sampling approach described in \citet{skilling_nested_2004} provides a reproducible, intuitive, and reliable method to ascertain how much more likely one hypothesis is compared to another. Using this approach, we can investigate the most likely representation of the microflare spectra.

\section{Comparing hypotheses} \label{comparing}

Bayes formula \citep{jaynes_probability_2003} is used extensively to test how well a certain hypothesis, $H$, explains given data, $D$, via the posterior, $P(H | D)$, and is given by
\begin{equation} \label{bayes-hyp}
    P(H | D) = \frac{P(D | H)P(H)}{P(D)},
\end{equation}
where $P(D | H)$ is termed the likelihood which describes how likely the data is observed if the hypothesis is true and $P(H)$ represents the prior information on the hypothesis \citep{sivia_data_2006}. The term $P(D)$ is the evidence which describes how likely the data is to occur independent of hypothesis which is difficult to obtain and, in some cases, has very little meaning. 

Different hypotheses for the given data can then be compared relative to each other using Equation~\ref{bayes-hyp} by calculating the posterior odds. To compute this between two different hypotheses---say $H_{1}$ and $H_{2}$---for the same data, the ratio is taken between the respective posteriors as 
\begin{equation} \label{post-odds}
    \frac{P(H_{1} | D)}{P(H_{2} | D)} = \frac{P(D | H_{1})}{P(D | H_{2})}\frac{P(H_{1})}{P(H_{2})},
\end{equation}
where we note the evidence term, $P(D)$, from Equation~\ref{bayes-hyp} cancels \citep{ashton_nested_2022}. Equation~\ref{post-odds}, therefore, provides a relative metric as to which hypothesis between $H_{1}$ and $H_{2}$ is more likely. It is common that the \textit{prior odds}, ratio of $P(H_{1})$ and $P(H_{2})$, is evaluated to $1$ if there is no prior justifiable preference for one hypothesis over another and so the posterior odds in this case is fully equivalent to the ratio of the likelihoods, termed \textit{Bayes factor} (BF). 

Using a nested sampling algorithm \citep{skilling_nested_2004,skilling_nested_2006}, we can estimate the likelihoods by investigating the parameter posterior under the assumption that the  hypothesis, now model $M$, is true. Nested sampling is a Bayesian tool used to compare parametric model fits to data where the models have parameters, $\theta$, and priors, $P(\theta | M)$. This is done by calculating the parameter posterior, $P(\theta | D, M)$, by 
\begin{equation} \label{mcmcBayes}
    P(\theta | D, M) = \frac{P(D | \theta, M)P(\theta | M)}{P(D | M)},
\end{equation}
where we note that all terms depend on the model, $M$, and the data, $D$, being fitted. This explicit form is typically presented in Bayesian analysis, such as in MCMC analysis \citep{ireland_estimating_2013} where the $P(D | M)$ term is often excluded in practice.

However, to avoid the repetition of nomenclature, we adopt another equivalent representation
\begin{equation} \label{bayes}
    P(\theta) = \frac{\mathcal{L}(\theta)\pi(\theta)}{\mathcal{Z}},
\end{equation}
where the nested sampling process uses the likelihood $\mathcal{L}(\theta)$ (e.g., Equation~\ref{likelihood}) and the prior $\pi(\theta)$ to estimate both the evidence $\mathcal{Z}$ and the posterior $P(\theta)$. 

Note that $\mathcal{Z}$ is equivalent to the likelihood term $P(D | H)$, in Equation~\ref{bayes-hyp} and~\ref{post-odds}, and is \textit{different} to the previous evidence term $P(D)$ from Equation~\ref{bayes-hyp}.

The evidence, $\mathcal{Z}$, could be calculated as 
\begin{equation} \label{fullEv}
    \mathcal{Z} = \int_{\theta}\mathcal{L}(\theta)\pi(\theta)d\theta ,
\end{equation}
ensuring the integral of $P(\theta)$ over the parameter-space is equal to~1 \citep{sivia_data_2006}. However, this integral can quickly become complex with an increasing number of parameter dimensions. The nested sampling process takes an approach to simplify this calculation. 

Estimating the evidences, $\mathcal{Z}$ (likelihood $P(D | H)$ in Equation~\ref{post-odds}) allows for a robust comparison between models and how well they explain the data. Criteria from \citet{kass_bayes_1995}, for example, can then be used to assign a relative confidence of one hypothesis compared to another. Importantly, this is also true even if the hypotheses rely on a different number of parameters or parameter priors. The next section details how nested sampling is performed and how $\mathcal{Z}$ is obtained mathematically.


\subsection{Calculating the evidence \texorpdfstring{$\mathcal{Z}$}, with nested sampling} \label{sec:maths} 

The evidence $\mathcal{Z}$ is calculated by sampling a predefined and physically motivated constrained prior, $\pi(\theta)$, for the model parameters while calculating the likelihood, $\mathcal{L}(\theta)$, at every corresponding location. From this sampling, a quantity called the \textit{prior mass} can be defined \citep{skilling_nested_2006}. The prior mass, $\xi$, is the amount of prior enclosed by some likelihood $\lambda$ which is given by
\begin{equation} \label{priorMass}
    \xi(\lambda) = \int_{\mathcal{L}(\theta)>\lambda}\pi(\theta)d\theta ,
\end{equation}
and contains all sampled points with $\mathcal{L}(\theta)>\lambda$. From Equation~\ref{priorMass}, we note that $\xi=1$ when $\lambda=0$ (all sampled points are enclosed) and so $\xi=0$ when $\lambda=\mathcal{L}_{max}$.

The definition of the prior mass can then simplify Equation~\ref{bayes} and~\ref{fullEv} to
\begin{equation} \label{bayesPM}
    P(\xi) = \frac{\mathcal{L}(\xi)}{\mathcal{Z}}
\end{equation}
and
\begin{equation} \label{fullEvPM}
    \mathcal{Z} = \int_{0}^{1}\mathcal{L}(\xi)d\xi ,
\end{equation}
respectively, where the evidence integral is now only performed over one dimension, the prior mass. Equation~\ref{bayesPM} shows taking a random sample of $\xi_{i}$ from~0--1, the range defined in Equation~\ref{fullEvPM}, provides a random sample from the posterior $P(\xi)$. This is equivalent to obtaining a random sample from the posterior $P(\theta)$ with a random parameter sample of $\theta_{i}$ as shown by Equation~\ref{bayes}. A nested sampling algorithm will then iterate through the random samples spread over the constrained prior to calculate the evidence and posterior distribution.


\subsubsection{The information $\mathcal{H}$} \label{sec:informationValue}

A useful value obtained throughout the nested sampling process is a quantity known as the information, $\mathcal{H}$. This quantity describes where the bulk of the posterior lies, with the posterior occupying a fraction of $e^{-\mathcal{H}}$ within the constrained prior \citep{skilling_nested_2006} and can also be used to estimate the likely number of iterations to compress the majority of the prior \citep{ashton_nested_2022}. For example, if the majority of the posterior takes up a small region of the prior then $\mathcal{H}$ is large and if most of the posterior occupies a large area of the prior then $\mathcal{H}$ is small.


\subsubsection{Iterating through the samples}

The Python nested sampling package, Nestle\footnote{\url{https://github.com/kbarbary/nestle}}, used throughout the analysis presented in Section~\ref{tornt} initially starts by sampling the constrained prior $n_{live}$ times where the corresponding likelihood value is calculated at each sample. These \textit{live-points} are then ordered from smallest likelihood, $\mathcal{L}_{*}$, to largest. The prior mass is then calculated for $\mathcal{L}_{*}$ before the corresponding live-point is removed and replaced with another sample with a likelihood value $>$$\mathcal{L}_{*}$ to conserve the number of samples across iterations. The prior mass and the likelihood of the removed live-point is recorded and used to build the integral described by Equation~\ref{fullEvPM}. Once a new live-point is chosen the process begins again and continues until the stopping criteria has been met \citep{skilling_nested_2004}.

The stopping criterion used in Nestle is one which estimates the remaining prior mass to be sampled. The iterative process can be stopped when the remaining estimated mass falls below a defined threshold, $A_{thresh}$, using
\begin{equation} \label{stopping}
    ln(\mathcal{L}_{max,i}\xi_{i} + \mathcal{Z}_{i}) - ln(\mathcal{Z}_{i}) < A_{thresh},
\end{equation}
where $\mathcal{L}_{max,i}$ is the maximum likelihood of the live-points, $\xi_{i}$ is the remaining prior mass, and $\mathcal{Z}_{i}$ is evidence value at iteration $i$. 

In addition to simplifying the evidence calculations, nested sampling also obtains many random samples of the posterior through the iterative procedure. Therefore, almost as a by-product, the same information that MCMC analysis provides is also obtained in the process. 

A nested sampling algorithm can be applied to many different problems involving model comparison. In Section~\ref{tornt} we apply the nested sampling algorithm used by Nestle to the active region X-ray microflare presented in Section~\ref{analysis} to determine the emission mechanisms present during its impulsive phase by testing different hypotheses for the observed data: the null, thermal, and non-thermal scenarios.


\section{Thermal or non-thermal emission} \label{tornt}

The analysis presented in Section~\ref{analysis} seems to indicate that non-thermal emission is present during this estimated A0.02 microflare's impulsive phase; however, the arguments are qualitative or weakly quantitative and far from decisive. Section~\ref{analysis-te} shows evidence of an earlier peaking and more impulsive time profile in the higher, 6--10~keV X-ray energies observed by NuSTAR compared to the lower energy microflare emission (Figure~\ref{fig:overview}, left panels). This supports the presence of an additional process beyond the one producing the EUV microflare signal, but it is not clear if this is due to a thermal or non-thermal source.

In section~\ref{analysis-spa}, we discussed the observed presence of a clear microflare loop structure in the \ion{Fe}{xviii} emission (Figure~\ref{fig:overview_sdo}), suggesting the presence of material between~4--10~MK which appears to be corroborated by all microflare spectral fits (Figure~\ref{fig:mcmc_spec}, right panel, and Figure~\ref{fig:mcmc_spec_fitExcess}). We also see loop footpoints across the SDO/AIA EUV channels that are anchored in two opposite magnetically polarised regions as viewed from SDO/HMI. Figure~\ref{fig:overview_sdo} indicates the presence of bright footpoints which could potentially be a result from heated material through microflare energised electrons decelerating in the lower solar atmosphere. This interpretation would align with the presence of non-thermal emission.

Performing spectral fitting analysis on the X-ray microflare spectra (Section~\ref{analysis-spe}) shows more evidence there is different behaviour below and above~6~keV. We find that an additional model is required to reasonably fit the count excess $>$6~keV beyond an isothermal fit and to resolve the poor fit $<$6~keV shown in Figure~\ref{fig:mcmc_spec} (right panel). However, it is still not possible to confidently determine if an additional thermal or non-thermal model is required (Figure~\ref{fig:mcmc_spec_fitExcess}). 

In order to obtain a robust measure of confidence over the nature of the high-energy~6--10~keV emission observed from the microflare, we make use of the nested sampling process (see Section~\ref{comparing}). The nested sampling result will then provide a vigorous, intuitive, and repeatable level of confidence on the emission mechanism most likely responsible for the higher energy X-ray emission.

\subsection{Nested sampling application to \evtid{}} \label{tornt-setup}

Since nested sampling relies on a constrained prior (Section~\ref{sec:maths}), effort must be taken to define appropriate bounds for each parameter. We make use of a uniform prior for all parameters where the bounds of each is determined by previous microflare studies with consideration to the thermal and non-thermal energetics involved.

\begin{table}
\centering
 \caption{The lower and upper limits to define the uniform constrained prior for each model parameter used for nested sampling. The same notation for each parameter used in the spectral fitting is given in brackets as well as the parameter units. The parameters T2 and EM2 are present in all microflare fits (Figure~\ref{fig:mcmc_spec} and~\ref{fig:mcmc_spec_fitExcess}) while T3 and EM3 are only in the thermal scenario fit and F$_{\text{e}^{-}}$, $\delta$, and E$_{c}$ are only in the non-thermal scenario fit shown in Figure~\ref{fig:mcmc_spec_fitExcess} to fit the count model excess.}
 \label{tab:priors}
 \begin{tabular}{lcccr}
  \hline
   Model Parameter & Lower    & Upper  \\
  \hline
  
   Microflare Temperature (T2, [MK]) &  2.5  &  10  \\[5pt]
   Microflare Emission Measure (EM2, [cm$^{-3}$]) & 1$\times$10$^{42}$ &  1$\times$10$^{46}$ \\[5pt]
          
    Excess Temperature (T3, [MK]) &  8  &  15 \\[5pt]
    Excess Emission Measure (EM3, [cm$^{-3}$]) & 1$\times$10$^{40}$ &  1$\times$10$^{44}$ \\[5pt]
          
    Excess Electron Flux (F$_{\text{e}^{-}}$, [e$^{-}$ s$^{-1}$]) &  1$\times$10$^{30}$  &  1$\times$10$^{33}$ \\[5pt]
    Excess Electron Spectral Index ($\delta$) & 4 &  15\\[5pt]

    Excess Low-energy Cut-off (E$_{c}$, [keV]) &  3  &  10     \\
  
  \hline
 \end{tabular}
\end{table}

Table~\ref{tab:priors} shows the bounds for the uniform priors used in the nested sampling analysis. Studies of microflares comparable in scale to the one studied here, estimated GOES class $\sim$A0.02, have suggested that those fitted with a single thermal model tend to have a temperature no higher than $\sim$10~MK with emission measure~$\lesssim$1$\times$10$^{46}$~cm$^{-3}$ \citep{duncan_nustar_2021,vadawale_observations_2021-1}. Additionally, some enhanced pre-flare emission may be present while some isothermal NuSTAR microflares have emission measures as low as $\sim$1$\times$10$^{43}$~cm$^{-3}$\citep{cooper_nustar_2021}. Therefore, sensible prior bounds for the microflare temperature and emission measure (T2 and EM2 from Figure~\ref{fig:mcmc_spec} and~\ref{fig:mcmc_spec_fitExcess}) are taken to be~2.5--10~MK and~1$\times$10$^{42}$--1$\times$10$^{46}$~cm$^{-3}$, respectively. 

From the physically motivated parameter prior ranges for T2 \& EM2, and using the volume estimate from Section~\ref{analysis-spa}, the possible thermal energies resulting from models within these bounds fall between approximately~4$\times$10$^{24}$--2$\times$10$^{27}$~erg. Considering past NuSTAR sub-A class microflares, this is a conservative thermal energy range and helps support the choice of the defined parameter prior bounds.

Next, priors must be defined for the thermal (T3 and EM3) and non-thermal (F$_{\text{e}^{-}}$, $\delta$, and E$_{c}$) excess model parameters. When other microflares have been found to be best fitted with two isothermal models, we find that the emission measure for the higher temperature component is $\sim$2 orders of magnitude less than the lower temperature model \citep{duncan_nustar_2021,cooper_nustar_2021} and so we choose a prior emission measure range of~1$\times$10$^{40}$--1$\times$10$^{44}$~cm$^{-3}$. In order to investigate a potential hotter temperature component producing the count model excess in Figure~\ref{fig:mcmc_spec} (right panel) we set a prior temperature range of~8--15~MK such that is overlaps with, and goes beyond, the range being investigated for T2.

The priors for the non-thermal excess models are determined from previous non-thermal X-ray microflare studies. \citet{hannah_rhessi_2008} finds~90\% of the RHESSI microflares studied have a low-energy cut-off between~9--16~keV with photon indices between~4--10 \cite[corresponding to electron indices between~5--11 for thick-target emission;][]{brown_deduction_1971}. However, the majority of these non-thermal X-ray microflares are of a greater GOES class and brighter than ones viewed by NuSTAR \citep{glesener_accelerated_2020,cooper_nustar_2021} which do not have emission extending to energies as high. Therefore, the conservative uniform prior bounds for the low-energy cut-off and the electron index are defined to be~3--10~keV and~4--15, respectively.

The microflare presented here is much weaker in magnitude than the non-thermal microflares investigated previously by NuSTAR and so the electron flux is expected to be less than previously obtained. Therefore, the prior range is estimated to be~10$^{30}$--10$^{33}$~electrons~s$^{-1}$ as this microflare is approximately three orders of magnitude weaker in GOES classification than the non-thermal microflare studied in \citet{glesener_accelerated_2020} which is found to have an electron flux of $\sim$2$\times$10$^{35}$~electrons~s$^{-1}$. The prior constraints for the non-thermal model provides an energy range $\sim$10$^{23}$--10$^{27}$~erg which is complementary to the thermal energy range being explored for the microflare providing additional physical justification for the thick-target parameter constrained prior bounds.

Throughout the nested sampling analysis, all three scenarios are run with~10,000 live-points and an $A_{thresh}$ value of~0.1. These, again, are conservative values to ensure the majority of the prior is sampled adequately. The nested sampling result does not change significantly with fewer live-points or a larger $A_{thresh}$, only the uncertainties become larger.

\subsection{Nested sampling results} \label{tornt-results}

Nested sampling analysis is performed on all three model representations of the microflare spectra where the process took on the order of hours to complete for each model scenario using an Apple M1 Max MacBook Pro. The thermal and non-thermal representations of the excess from Figure~\ref{fig:mcmc_spec_fitExcess} (left and right panel) give $ln(\mathcal{Z})$ values of $-497.31\pm0.03$ and $-493.63\pm0.03$, respectively, while the null scenario, where the excess is not represented (Figure~\ref{fig:mcmc_spec}, right panel), is estimated to have a value of $-530.14\pm0.03$. 

\begin{table}
\centering
 \caption{The logarithm of Bayes factor, Bayes factor, and the corresponding level of confidence from \citet{kass_bayes_1995} are shown for each model scenario comparison for the microflare spectra. The error on $ln$(BF) is~$\pm$0.04 for all values.}
 \label{tab:BFs}
 \begin{tabular}{lccccr}
  \hline
   Comparison & $ln$(BF) &  BF  &  Confidence \\
  \hline
  
   $\mathcal{Z}_{thermal}$-to-$\mathcal{Z}_{null}$ & 32.8 &  2$\times$10$^{14}$  &  Very Strong  \\
   $\mathcal{Z}_{non-thermal}$-to-$\mathcal{Z}_{null}$ & 36.5 &  7$\times$10$^{15}$  &  Very Strong  \\
   $\mathcal{Z}_{non-thermal}$-to-$\mathcal{Z}_{thermal}$ & 3.7 &  40  &  Strong  \\
  
  \hline
 \end{tabular}
\end{table}

The $\mathcal{H}$ values (as described in Section~\ref{sec:informationValue}) reported for the analysis in each case is given to be $\sim$9 for the null scenario and $\sim$11 for both the thermal and non-thermal cases. The similar values indicate that the bulk of the posterior lies within a similar fraction of the constrained prior for each tested hypothesis.

\begin{figure*}
        \includegraphics[width=0.55\linewidth]{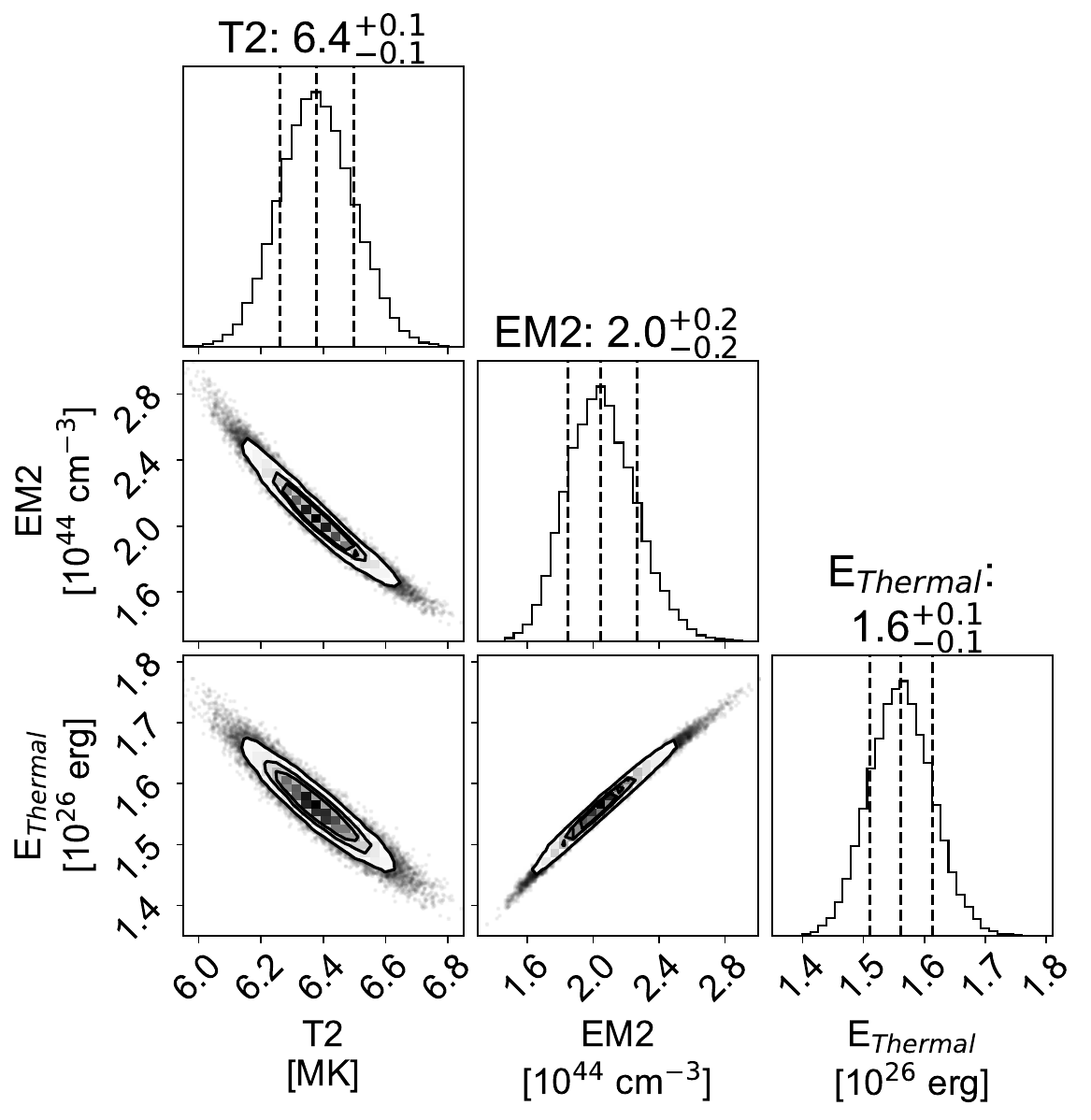}
    \caption{Posterior distributions for the isothermal, \textit{null scenario}, representation of the microflare spectra shown in Figure~\ref{fig:mcmc_spec} (right panel) as determined from nested sampling analysis. The thermal energy distribution is calculated from the temperature (T2) and emission measure (EM2) samples. The median and 1-$\sigma$ confidence intervals are displayed above the diagonal panels for each marginalised parameter distribution and shown with dashed lines. Contours of 50\%, 68\%, and 90\% levels are shown in the 2D marginalised posterior distribution.}
    \label{fig:nest_null_corner}
\end{figure*}

\begin{figure*}
        \includegraphics[width=0.8\linewidth]{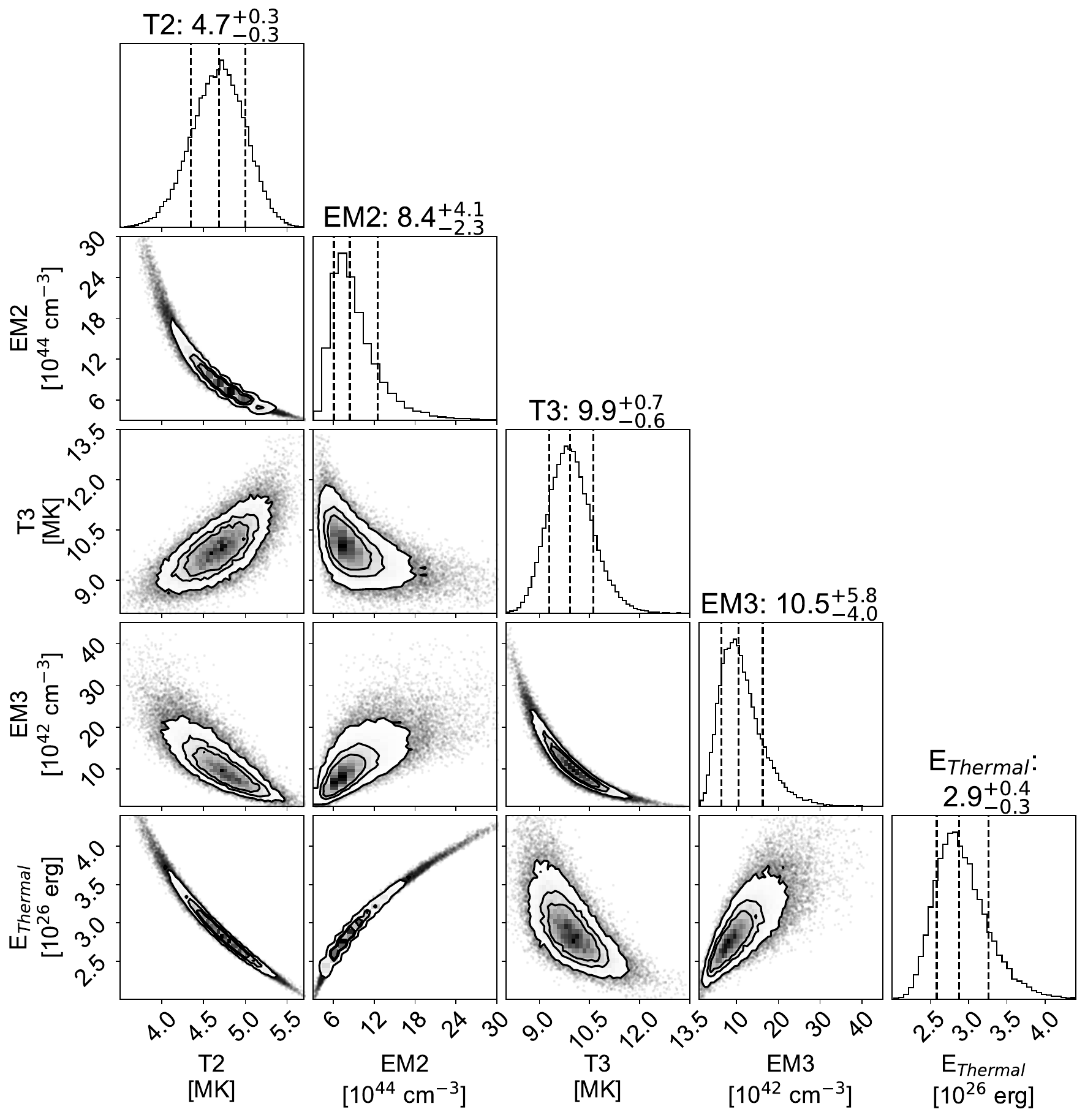}
    \caption{Similar to Figure~\ref{fig:nest_null_corner} but the nested sampling analysis result for the \textit{thermal scenario} shown in the left panel of Figure~\ref{fig:mcmc_spec_fitExcess}. The thermal energy distribution is calculated from the temperature (T2 \& T3) and emission measure (EM2 \& EM3) samples.}
    \label{fig:nest_thermal_corner}
\end{figure*}

\begin{figure*}
        \includegraphics[width=1\linewidth]{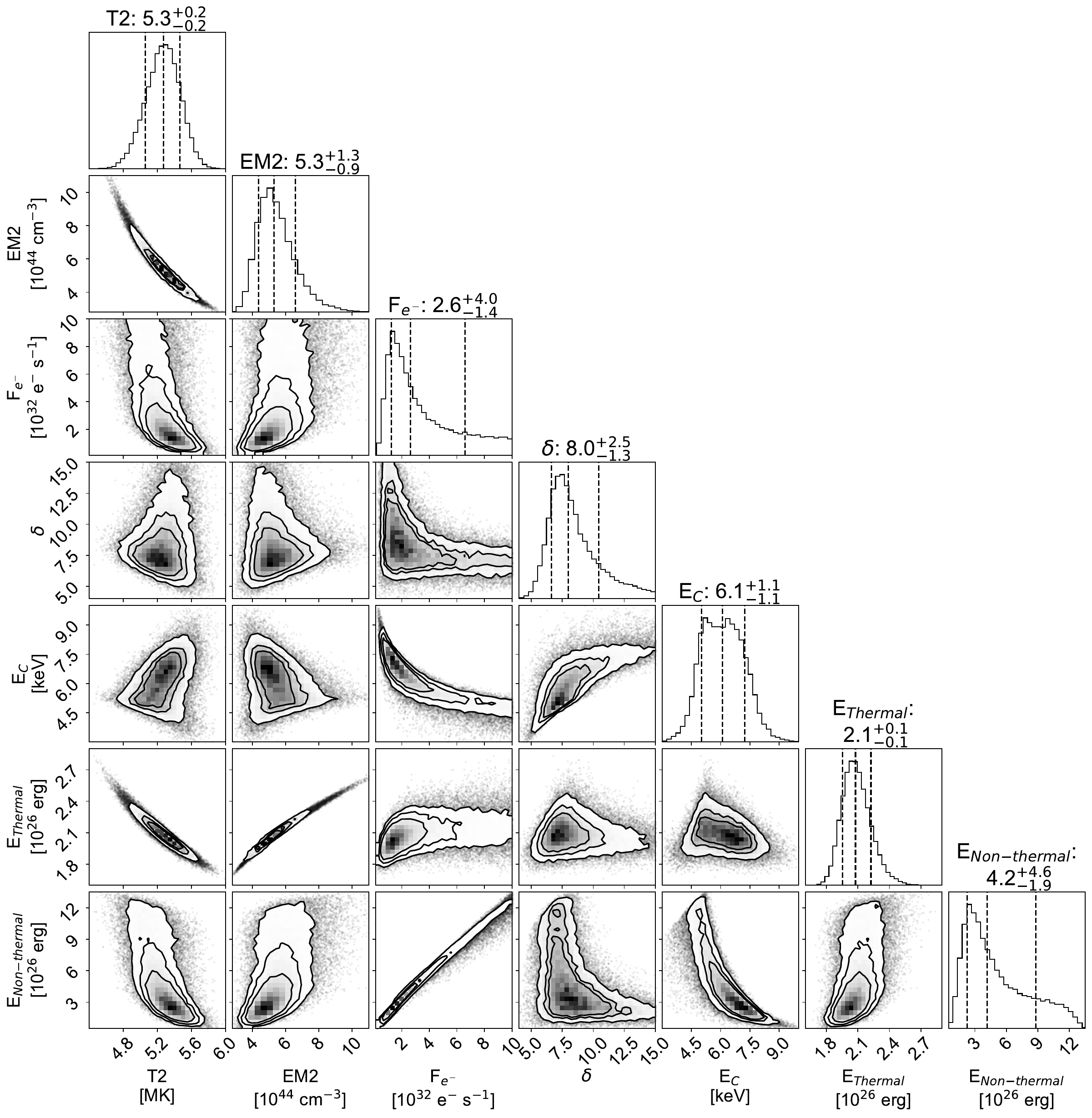}
    \caption{Similar to Figure~\ref{fig:nest_null_corner} but the nested sampling analysis result for the \textit{non-thermal scenario} shown in the right panel of Figure~\ref{fig:mcmc_spec_fitExcess}. The thermal energy distribution is calculated from the temperature (T2) and emission measure (EM2) samples and the non-thermal energy distribution is calculated from the electron flux (F$_{\text{e}^{-}}$), electron spectral index ($\delta$), and the low-energy cut-off (E$_{c}$) samples.}
    \label{fig:nest_nonthermal_corner}
\end{figure*}

The posterior distribution corner plots that accompany the evidences are shown in Figure~\ref{fig:nest_null_corner},~\ref{fig:nest_thermal_corner}, and ~\ref{fig:nest_nonthermal_corner}. We find that the constructed corner plots from the nested sampling analysis agree closely with those produced from the previous MCMC analysis performed in Section~\ref{analysis-spe} with all parameter values agreeing well within the 1-$\sigma$ equivalent uncertainties. This provides confidence that the posterior distribution for each model parameter, bound by the defined priors, is sampled well in the MCMC and that there are no other relevant solutions inside the physically motivated prior bounds.

With the confirmation from the corner plots produced from the nested sampling analysis, and corroboration with the MCMC analysis that the evidences have been calculated robustly, we are able to compare each model representation to the others using Bayes factor from Equation~\ref{post-odds}. This is reasonable as we have no evidence or reason to prefer one model representation over the other; therefore, we assume a prior odds ratio of~1 making the BF equal to the posterior odds.

The BF for each comparison is shown in Table~\ref{tab:BFs}. We find that the BF comparing the thermal and non-thermal representations of the count excess are far more probable than not representing the excess at all in the null scenario. Therefore, nested sampling has given a quantitative measure of how much an additional model is required beyond the pre-flare and isothermal fit with the thermal and non-thermal scenarios being~$>$10$^{14}$ times more probable than the null hypothesis.

Additionally, the nested sampling also shows that the non-thermal representation is $\sim$40 times more likely to be the cause of the count excess compared to the thermal model, corresponding to strong evidence for this conclusion. Therefore, the nested sampling analysis would suggest that this event is the weakest X-ray microflare to show direct evidence for non-thermal emission.

To check the sensitivity of the nested sampling result on the choice of physically motivated prior bounds from Table~\ref{tab:priors}, we adjust the bounds of each parameter prior by up to~20\%. This still results in the non-thermal scenario being more preferable to the thermal representation with $ln$(BF) values between~3.5--3.8.

Furthermore, we narrow the priors around the bulk of the T3 and EM3 distributions shown in Figure~\ref{fig:nest_thermal_corner} while leaving the non-thermal parameter priors unchanged in order to be more favourable to the thermal scenario. An $ln$(BF) value of~$\sim$2 is obtained from this test, which is still classified as positive evidence for the non-thermal representation of the microflare \citep{kass_bayes_1995}.

We note that the best-fit log-likelihood values, $ln(\mathcal{L}_{opt})$, stated in Section~\ref{sec:best-spec-fit} show similar differences to the log-evidence values, $ln(\mathcal{Z})$, for the different scenarios; however, the nested sampling result is more informative and robust for model comparison. The evidences can sufficiently compare models with different parameters and priors, whereas a clear conclusion cannot be drawn from the best-fit log-likelihood values as discussed in Section~\ref{sec:best-spec-fit} and~\ref{comparing}.

\section{Summary and conclusions}

In this paper, we present the first analysis of X-ray flare spectra using a nested sampling algorithm. This microflare is estimated to be approximately an A0.02 GOES equivalent class flare which shows signs of high temperatures and/or non-thermal particle acceleration; however, determining the extent of this emission is difficult using arguments and methods performed in the past.

From the temporal, spatial, and spectral information discussed in Section~\ref{analysis}, we find evidence that the X-ray microflare emission is not well represented with an isothermal model as a count excess is present at higher energies in the NuSTAR spectra (Figure~\ref{fig:mcmc_spec}). Upon investigating two emission mechanisms potentially responsible for this we find that both a thermal and non-thermal explanation for the excess appear to be equally valid with no decisive reason to pick one over the other.

Qualitative evidence for a non-thermal representation is found in the form of the lack of higher temperature ($\sim$10~MK) emission response in the~131~\AA{} SDO/AIA channel (Figure~\ref{analysis-te}) and potentially in the presence of bright EUV loop footpoints visible throughout the solar atmosphere (Figure~\ref{analysis-spa}). When performing spectral fitting, the model with the non-thermal component produces a slightly more preferable log-likelihood value (Figure~\ref{fig:mcmc_spec_fitExcess}); however, this metric may be expected to be better due to the non-thermal component having an extra free parameter to use during the fitting process compared to the thermal model. Therefore, a conclusion cannot be made based on the best-fit log-likelihood values.

We use a more robust model comparison method called nested sampling (Section~\ref{comparing}) which is used to determine an estimate for the evidence attributed to each model representation of the data, therefore, allowing Bayes factor to be computed for model comparison. In the process we also obtain the posterior distributions for each model as shown in Figure~\ref{fig:nest_null_corner},~\ref{fig:nest_thermal_corner}, and~\ref{fig:nest_nonthermal_corner}. We find that the model containing the non-thermal component is most likely to represent the observed microflare spectra with a BF of 7$\times$10$^{15}$ and~40 compared to the null and thermal representation, respectively.

Therefore, we report the weakest X-ray microflare to have direct observation of non-thermal particle emission suggesting that X-ray flares of the estimated scale A0.02 do indeed show similar traits to their larger counterparts. From the nested sampling results, this microflare produces a thermal energy of~2.1$^{+0.1}_{-0.1}\times$10$^{26}$~erg which is rational compared to the non-thermal energy of~4.2$^{+4.6}_{-1.9}\times$10$^{26}$~erg released over the course of its impulsive phase at 3.0$^{+3.3}_{-1.4}\times$10$^{24}$~erg~s$^{-1}$. The corresponding energy posterior distributions of each model scenario are shown in Figure~\ref{fig:nest_null_corner},~\ref{fig:nest_thermal_corner}, and~\ref{fig:nest_nonthermal_corner}. This shows that microflares this weak, and likely weaker, continue to undergo the same physical processes with the same/similar emission mechanisms as brighter flares.

The nested sampling algorithm is in no way specific to X-ray microflare analysis. It can use used to gain insight into general model comparison problems and is able to produce the same data products as MCMC analysis. However, effort will be made to speed up the process specific to this microflare analysis. This could be achieved by speeding up the code used for the individual component models and testing a range of nested sampling algorithm implementations such as diffusive nested sampling \citep{brewer_diffusive_2011} and those discussed in \citet{speagle_dynesty_2020} and \citet{williams_importance_2023}. 

Using nested sampling has enabled the robust analysis of the high-energy count excess. Methods to determine the emission mechanism for such features in the past are either primarily subjective in nature and/or fail to give a reliable conclusion. The nested sampling algorithm is able to allow the repeatable and intuitive study of data with low signal and is crucial when investigating microflares at such weak scales. Nested sampling will be used in future NuSTAR microflare studies, and other flare studies with ambiguous spectra, to determine the confidence on the emission mechanisms present throughout the event.

\section*{Acknowledgements}


This paper made use of data from the NuSTAR mission, a project led by the California Institute of Technology, managed by the Jet Propulsion Laboratory, funded by the National Aeronautics and Space Administration. These observations were supported through the NuSTAR Guest Observer program (NASA grant 80NSSC21K0135). This research used version 4.1.5 \citep{stuart_j_mumford_2021_4421322} of the SunPy open source software package \citep{sunpy_community_sunpy_2020}, version 0.7.3 \citep{barnes_w_t_2020_4315741} of the Aiapy open source software package \citep{barnes_aiapy_2020}, and made use of Astropy,\footnote{\url{http://www.astropy.org}} a community-developed core Python package for Astronomy \citep{2018AJ....156..123T}. Other Python packages that were extensively used were Matplotlib \citep{Hunter:2007}, NumPy \citep{harris2020array}, and SciPy \citep{2020SciPy-NMeth}. This research also made use of HEASoft (a unified release of FTOOLS and XANADU software packages) and NuSTAR Data Analysis Software (NuSTARDAS). This paper also made use of the SolarSoft IDL distribution (SSW) from the IDL Astronomy Library.

KC was supported by a Royal Society Research Fellows Enhancement Award RGF$\backslash$EA$\backslash$180010 and IGH was supported by STFC grant ST/T000422/1. This work is also supported by the NSF career grant AGS 1752268. The authors would like to thank John Veitch and Christopher Messenger for their insight and helpful discussions with regards to the nested sampling process. We would also like to thank the anonymous referee for their helpful feedback.

\section*{Data Availability}


All data used is publicly available. SDO data can be obtained from the Joint Science Operations Center (JSOC)\footnote{\url{http://jsoc.stanford.edu/}} using SunPy's \verb|Fido|\footnote{\url{https://docs.sunpy.org/en/stable/guide/acquiring_data/fido.html\#fido-guide}} object while the NuSTAR data is available from the NuSTAR Master Catalog\footnote{\url{https://heasarc.gsfc.nasa.gov/db-perl/W3Browse/w3table.pl?tablehead=name=numaster&Action=More+Options}} with the OBSID 20618003001. An overview of NuSTAR solar observation campaigns is also available on Github.\footnote{\url{https://ianan.github.io/nsigh_all/}}



\bibliographystyle{mnras}
\bibliography{refs}




\appendix

\section{X-ray thermal and non-thermal photon models} \label{app:ph_mods}

\begin{figure*}
        \includegraphics[width=1\linewidth]{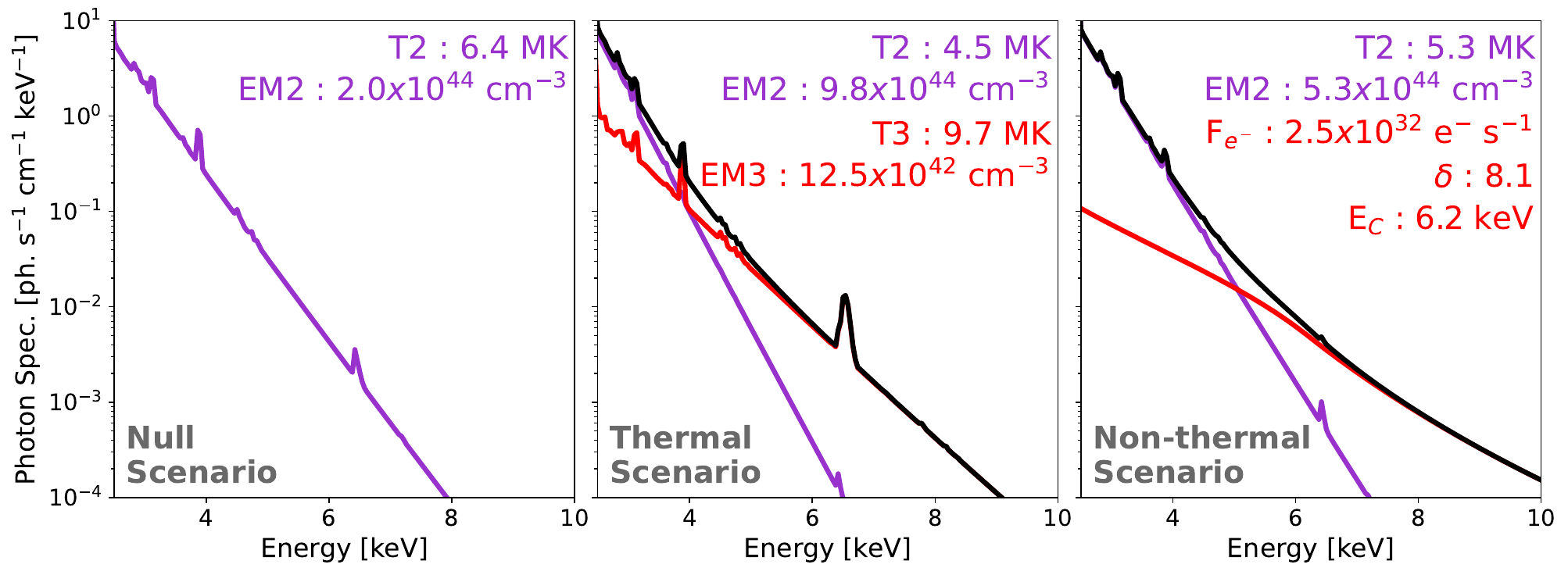}
    \caption{Photon models, obtained from \sunxspex{}, used to fit the NuSTAR microflare X-ray count spectra in the \textit{null} (left panel), \textit{thermal} (middle panel), and \textit{non-thermal} (right panel) \textit{scenarios} shown in Figure~\ref{fig:mcmc_spec}'s right panel and Figure~\ref{fig:mcmc_spec_fitExcess}. The models shown are calculated using NuSTAR's native energy resolution and the parameter values obtained from MCMC analysis (Section~\ref{analysis-spe}). Each model component, and corresponding parameter set, is shown in purple or red with the total model in the thermal and non-thermal case being shown in black.}
    \label{fig:photon_models}
\end{figure*}

Figure~\ref{fig:photon_models} displays the photon models used to investigate the null (left panel), thermal (middle panel), and non-thermal (right panel) scenarios for the microflare as initially described in Section~\ref{analysis-spe}. Two photon models are utilised to represent each scenario, an isothermal model and a thick-target non-thermal model. Both models represent a photon spectrum produced from different electron distributions.

The isothermal model (purple in all panels and red in the middle panel of Figure~\ref{fig:photon_models}) represents a photon spectrum produced from a Maxwell-Boltzmann distribution of electrons. The isothermal model is composed of continuum \& line emission and is a function of the plasma temperature (T) and emission measure (EM).

The non-thermal photon spectrum (red in the right panel of Figure~\ref{fig:photon_models}) results from a power-law electron distribution losing all their energy through Coulomb collisions \citep{brown_deduction_1971,holman_implications_2011}. The thick-target model is a function of electron flux (F$_{\text{e}^{-}}$), electron spectral index ($\delta$), and low-energy cut-off (E$_{c}$).

\bsp	
\label{lastpage}
\end{document}